\documentclass[aps,pra,twocolumn,amsmath,amssymb,nofootinbib,showpacs,superscriptaddress]{revtex4}
\usepackage[english]{babel}
\usepackage{latexsym}
\usepackage{graphics}
\usepackage{epsfig}
\usepackage{color}
\usepackage{bm}
\usepackage{amsmath}
\usepackage{amssymb}

\usepackage[normalem]{ulem}
\usepackage{color}

\newcommand{\1}{\uparrow}
\newcommand{\2}{\downarrow}

\begin{document}

\title{Dimer-dimer zero crossing and dilute dimerized liquid in a one-dimensional mixture}
\author{A.~Pricoupenko}
\affiliation{LPTMS, CNRS, Univ. Paris Sud, Universit\'e Paris-Saclay, 91405 Orsay, France}
\affiliation{D\'epartement de Physique, \'Ecole Normale Sup\'erieure de Lyon, 46 All\'ee d'Italie, 69007 Lyon}
\author{D.~S.~Petrov}
\affiliation{LPTMS, CNRS, Univ. Paris Sud, Universit\'e Paris-Saclay, 91405 Orsay, France}

\date{\today}

\begin{abstract}

We consider the system of dimers formed in a one-dimensional mass-balanced Bose-Bose mixture { of species $\sigma=\1,\2$} with attractive interspecies and repulsive intraspecies contact interactions. In the plane parametrized by the ratios of the coupling constants $g_{\1\1}/|g_{\1\2}|$ and $g_{\2\2}/|g_{\1\2}|$ we trace out the curve where the dimer-dimer interaction switches from attractive to repulsive. We find this curve to be significantly (by more than a factor of 2) shifted towards larger $g_{\sigma\sigma}$ (or smaller $|g_{\1\2}|$) compared to the mean-field stability boundary $g_{\1\1}g_{\2\2}=g_{\1\2}^2$. For a weak dimer-dimer attraction we predict a dilute dimerized liquid phase stabilized against collapse by a repulsive three-dimer force.

\end{abstract}

%\pacs{34.50.-s, 05.30.Jp, 67.85.-d}

\maketitle

\section{Introduction}

Bosonic mixtures with competing attractive interspecies and repulsive intraspecies interactions have recently caught a great deal of attention because of their ability to form { droplets which exist without confinement and exhibit peculiar quantum properties} \cite{Petrov,Tarruell1,Fattori,Tarruell2}. Three-dimensional mixtures liquefy for $|g_{\1\2}|>\sqrt{g_{\1\1}g_{\2\2}}$, i.e., in the regime where the mean-field theory predicts collapse (we only consider $g_{\1\2}<0$) \cite{Petrov}. The stabilization comes from an effectively repulsive beyond-mean-field Lee-Huang-Yang term. By contrast, in one dimension this beyond-mean-field correction is effectively {\it attractive}; in the thermodynamic limit the mixture collapses for $|g_{\1\2}|>\sqrt{g_{\1\1}g_{\2\2}}$ and liquefies for $|g_{\1\2}|<\sqrt{g_{\1\1}g_{\2\2}}$ \cite{PetrovAstrakharchik}. 

These results are valid in the weakly-interacting regime close to the collapse line $|g_{\1\2}|=\sqrt{g_{\1\1}g_{\2\2}}$ where the saturation density of the liquid is high. Departing from this line (by increasing $g_{\sigma\sigma}$ or decreasing $|g_{\1\2}|$) makes the system more dilute, which, in one dimension, leads to stronger correlations. For $g_{\sigma\sigma}\gg |g_{\1\2}|$ the one-dimensional mixture with equal $\1$ and $\2$ populations eventually becomes a gas of $\1\2$ dimers with interdimer repulsion. Indeed, in the limit $g_{\1\1}=g_{\2\2}=\infty$ the two bosonic components can individually be mapped to noninteracting fermions \cite{Girardeau1,Girardeau2} and their mixture becomes equivalent to the exactly solvable fermionic Gaudin-Yang model \cite{Gaudin,Yang,review} which has no bound states other than $\1\2$ dimers.

In this paper we make a step towards understanding the nonperturbative intermediate region by starting from the repulsive gas of dimers and decreasing the ratio $g_{\1\1}g_{\2\2}/g_{\1\2}^2$ for finite generally different intraspecies coupling constants. We calculate the curve in the plane $\{g_{\1\1}/|g_{\1\2}|,g_{\2\2}/|g_{\1\2}|\}$ where the dimer-dimer interaction vanishes. In its vicinity we find the dimer-dimer scattering length and effective range.
We also show that collisions between dimers in this regime do not lead to the formation of trimers. On the attractive side of the dimer-dimer zero crossing we predict the existence of a dilute liquid of dimers stabilized against collapse by a repulsive three-dimer interaction. We describe properties of this state by developing a one-dimensional version of the mean-field theory of droplets proposed by Bulgac \cite{Bulgac}. By using an appropriate wave-function mapping, our results also apply to one-dimensional Bose-Fermi mixtures.  

The paper is organized as follows. In Sec.~\ref{Sec:Formalism} we derive a set of integral equations for solving the general $N$-body problem with zero-range interactions in one dimension. In Sec.~\ref{Sec:Trimer} we calculate the $\1\1\2$ trimer binding energy and thus determine the trimer-formation threshold in dimer-dimer collisions. In Sec.~\ref{Sec:Four-body} we find the dimer-dimer scattering parameters in the regime of a nearly vanishing effective dimer-dimer interaction. In Sec.~\ref{Sec:Liquid} we apply the mean-field theory of Bulgac to the dilute system of dimers taking care of the one-dimensional three-dimer interaction. In Sec.~\ref{Sec:Discussion} we discuss other possible scenarios for the mixture and present an outlook for further studies.

\section{One-dimensional $N$-body integral equation}\label{Sec:Formalism}

Direct solution of the Schr\"odinger equation may not be the most efficient way of solving the problem of particles interacting via zero-range potentials. Skorniakov and Ter-Martirosian (STM), by using the zero-range approximation for internucleon forces, reduced the three-body problem of neutron-deuteron scattering to a one-dimensional integral equation \cite{STM}. The STM approach has proved its power for many problems with ultracold atoms where the interactions can indeed with a very good accuracy be considered zero range. A great advantage of the method is that it works directly in the zero-range limit using two-body scattering parameters (scattering length, effective range) as the starting point. This allows one to concentrate on few- and many-body processes, bypassing the task of solving the two-body scattering problem on the way. The STM equation in a very general form for $N$ particles interacting via zero-range potentials in any dimension is derived in Ref.~\cite{Pricoupenko}. Here we present a self-contained derivation in the one-dimensional case keeping arbitrary $N$, masses, and number of species for future reference.

The Schr\"odinger equation for $N$ particles of masses $m_i$ { moving in free space and} interacting via zero-range potentials with coupling constants $g_{ij}$ reads
\begin{widetext}
\begin{equation}\label{SchrCoord} 
\left[-\sum_{i=1}^{N}\frac{1}{2m_i}\frac{\partial^2}{\partial x_i^2}-E\right]\psi(x_1,...,x_N)=-\sum_{i<j}g_{ij}\delta(x_i-x_j)\psi(x_1,...,x_N),
\end{equation}
where $E$ is the energy and we set $\hbar=1$. Introducing the Fourier transform $\psi(p_1,...,p_{N})=\int e^{{-ip_1 x_1...-ip_N x_N}}\psi(x_1,...,x_N)dx_1...dx_N$, switching to momentum space, and restricting our analysis to negative energies $E<0$ we rewrite Eq.~(\ref{SchrCoord}) in the form
\begin{equation}\label{SchrMom} 
\psi(p_1,...,p_{N})=-\frac{\sum_{i<j}g_{ij}{{\cal F}}_{ij}(p_1,...,p_{i-1},p_{i+1},...,p_{j-1},p_{j+1},...,p_{N}{;Q})}{\sum_{i=1}^{N}p_i^2/2m_i-E},
\end{equation}
where { $Q=\sum_{i=1}^N p_i$ and ${\cal F}_{ij}$ is the Fourier transform of $\delta(x_i-x_j)\psi(x_1,...,x_N)$ or, alternatively,
\begin{equation}\label{F}
{\cal F}_{ij}(p_1,...,p_{i-1},p_{i+1},...,p_{j-1},p_{j+1},...,p_{N};Q)=\int \psi(p_1,...,{p_i}',...,{p_j}',...,p_N)2\pi\delta({p_i}'+{p_j}'-p_i-p_j)\frac{d{p_i}' d{p_j}'}{(2\pi)^2}.
\end{equation} 
The center-of-mass momentum $Q$ is a conserved parameter and without loss of generality we take ${\cal F}_{ij}(p_1,...,p_{i-1},p_{i+1},...,p_{j-1},p_{j+1},...,p_{N}; Q)=2\pi \delta(Q)F_{ij}(p_1,...,p_{i-1},p_{i+1},...,p_{j-1},p_{j+1},...,p_{N})$}. We can now substitute Eq.~(\ref{SchrMom}) into Eq.~(\ref{F}). This eliminates $\psi$ and straightforwardly leads to the STM equations
\begin{equation}\label{STMGen} 
F_{ij}(p_1,...,p_{i-1},p_{i+1},...,p_{j-1},p_{j+1},...,p_{N})=-\int\frac{\sum_{k<l}g_{kl}F_{kl}(p_1,...,p_{k-1},p_{k+1},...,p_{l-1},p_{l+1},...,p_{N})}{\sum_{k=1}^{N}p_k^2/2m_k-E}\delta\left(\sum_{k=1}^N p_k\right)\frac{dp_i dp_j}{2\pi},
\end{equation} 
\end{widetext}
One can think of the function $F_{ij}$ as a wave function for $N-2$ atoms plus a pair { with momentum opposite to the total momentum of the atoms. Therefore, there are only $N-2$ arguments in $F$. The function} $F_{ij}$ has the same symmetry with respect to permutation of its arguments as the total wave function $\psi$. When these symmetries are taken into account, the number of different functions $F_{ij}$ needed to describe the system reduces to the number of coupling constants characterizing different interactions in the mixture. 

In the two-body case the function $F_{12}$ is a number and the STM equation is a simple algebraic equation for the determination of the dimer binding energy, $\sqrt{-2E/\mu_{12}}=-g_{12}$, where $\mu_{12}=m_1m_2/(m_1+m_2)$. 

\section{Trimer energy}\label{Sec:Trimer}

Before we embark on the dimer-dimer scattering let us make a brief detour into the three-body problem. We need to know the trimer binding energies in order to determine the trimer-formation threshold in dimer-dimer collisions. In the considered case of repulsive intraspecies couplings three identical bosons do not bind. Below, we analyze the $\1\1\2$ combination, the results being obviously valid for the $\1\2\2$ system upon interchanging $g_{\1\1}$ and $g_{\2\2}$ (we consider $m_\1=m_\2$). The $\1\1\2$ trimer can be formed if $\epsilon_{\1\1\2}$ is smaller than $E$, which, for zero dimer-dimer collision energy, equals twice the $\1\2$ dimer energy.

Consider the $\1\1\2$ equal-mass bosonic problem characterized by the coupling constants $g_{\1\1}=-2/a_{\1\1}$ and $g_{\1\2}=-2/a_{\1\2}<0$, where $a_{\sigma\sigma'}$ are the one-dimensional scattering lengths and we set $m_\1=m_\2=1$. Let us define $F_{\1\1}(p)=F_{12}(p)$ and $F_{\1\2}(p)=F_{13}(p)=F_{23}(p)$, the latter equality follows from Eq.~(\ref{F}) and the symmetry $\psi(p_1,p_2,p_3)=\psi(p_2,p_1,p_3)$. The STM equations read
\begin{equation}\label{STM3body}
\begin{aligned}
&\!\!\!\left(\!1+\frac{g_{\1\1}}{\sqrt{3p^2-4E}}\right)\!F_{\1\1}(p)=\!\int\frac{2g_{\1\2}F_{\1\2}(q)}{E-p^2-pq-q^2}\frac{dq}{2\pi},\\
& \hspace{0cm} 
\!\!\!\left(\!1+\frac{g_{\1\2}}{\sqrt{3p^2-4E}}\right)\!F_{\1\2}(p)=\!\int\frac{g_{\1\2}F_{\1\2}(q)+g_{\1\1}F_{\1\1}(q)}{E-p^2-pq-q^2}\frac{dq}{2\pi}.
\end{aligned}
\end{equation}
In Fig.~\ref{Fig:Trimer} we plot the trimer energy $\epsilon_{\1\1\2}<0$ in units of the dimer binding energy $|\epsilon_{\1\2}|=1/a_{\1\2}^2$ as a function of the ratio $g_{\1\1}/|g_{\1\2}|$. The curve is obtained by { discretizing the momentum, transforming the integrals in Eqs.~(\ref{STM3body}) into sums, and solving the resulting matrix-eigenvalue problem}. %solving a discretized version of Eqs.~(\ref{STM3body}) \textcolor{red}{i.e. by putting equations on a grid and transform the integrals into sums such that we end up with a matrix problem.} 
 Also shown are the atom-dimer (dotted) and dimer-dimer (dashed) scattering thresholds.

For the considered case of negative $g_{\1\2}$ the $\1\1\2$ trimer is always bound. However, in the limit $g_{\1\1}=\infty$ the trimer binding energy $\epsilon_{\1\1\2}-\epsilon_{\1\2}$ vanishes and the atom-dimer even-channel scattering length diverges. This limit is a particular case of the exactly solvable $N+1$ McGuire \cite{McGuire1965,McGuire1966} or more general Gaudin-Yang model \cite{Gaudin,Yang} of attractive spin-$1/2$ fermions. The connection with our bosonic system is obtained by the wave-function mapping $\psi_{\rm Bose}(x_{\1 1},x_{\1 2},x_\2)=\psi_{\rm Fermi}(x_{\1 1},x_{\1 2},x_\2){\rm sign}(x_{\1 1}-x_{\1 2})$ \cite{Girardeau1,Girardeau2}. In the fermionic case it is thus the odd-channel atom-dimer scattering length that diverges. The explicit expression for the atom-dimer transmission amplitude (there is no reflection) in this case is given in Ref.~\cite{Sutherland}.

The trimer state deepens with decreasing $g_{\1\1}/|g_{\1\2}|$. We find numerically that it crosses the dimer-dimer threshold, i.e., its energy equals $\epsilon_{\1\1\2}=2\epsilon_{\1\2}$, for $g_{\1\1}=0.0738|g_{\1\2}|$. It is worth mentioning that for  $g_{\1\1}=-|g_{\1\2}|$ the ground state of our $\1\1\2$-system is the same as the ground state of three identical attractive bosons. The trimer energy here equals four times the dimer energy \cite{McGuireBosons}. 

\begin{center}
\begin{figure}[ht]
\vskip 0 pt \includegraphics[clip,width=0.9\columnwidth]{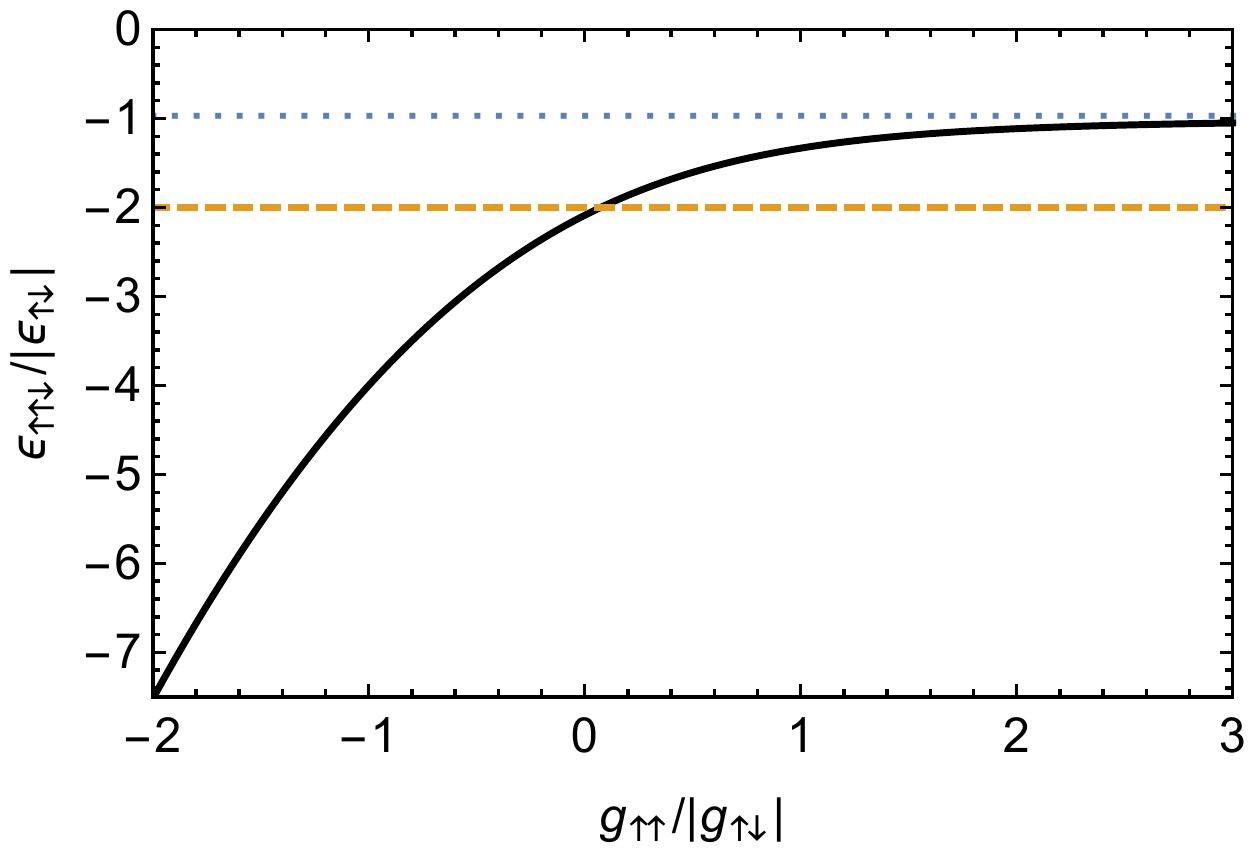}
\caption{
The $\1\1\2$-trimer energy in units of the $\1\2$-dimer energy as a function of $g_{\1\1}/|g_{\1\2}|$ (solid). The dotted and dashed lines indicate, respectively, the atom-dimer and dimer-dimer scattering thresholds.      
}
\label{Fig:Trimer}
\end{figure}
\par\end{center}
 
\section{Dimer-dimer scattering problem}\label{Sec:Four-body}  

Consider now the scattering problem of two $\1\2$ dimers and let $1$ and $2$ refer to $\1$ particles and $3$ and $4$ -- to $\2$ particles. Then, from Eq.~(\ref{F}) and from the { symmetry relations} $\psi(p_1,p_2,p_3,p_4)=\psi(p_2,p_1,p_3,p_4)=\psi(p_1,p_2,p_4,p_3) { = \psi(p_2,p_1,p_4,p_3)}$, one can show that $F_{13}=F_{14}=F_{23}=F_{24}$. We thus denote this function by $F_{\1\2}$ and also define $F_{\1\1}=F_{12}$ and $F_{\2\2}=F_{34}$. Equations~(\ref{STMGen}) then transform into a set of three coupled two-dimensional homogeneous equations for $F_{\sigma\sigma'}$. We do not write these rather bulky equations to avoid cluttering. However, they are well suitable for numerical calculations on a grid. Let us just explain how we deduce the dimer-dimer scattering amplitude from the numerics.

We look for the dimer-dimer scattering solution at the collision energy $p_0^2/2$, which corresponds to $E=-2|\epsilon_{\1\2}|+p_0^2/2$. In real space, when the two dimers are separated from each other by more than their size, $|x_1+x_3-x_2-x_4|/2\gg a_{\1\2}$, the four-body wave function factorizes into 
\begin{equation}\label{PsiAsymp}
\psi(x_1,x_2,x_3,x_4) \approx \phi_{0}(x_{13}) \phi_{0}(x_{24}) \chi[(x_1+x_3-x_2-x_4)/2],
\end{equation}
where $x_{ij}=x_i-x_j$, $\phi_0(r)=\sqrt{1/a_{\1\2}}\exp(-|r|/a_{\1\2})$ is the normalized dimer wave function, $\chi(R)=\cos(p_0 R)+f(p_0)\exp(ip_0|R|)$ describes the relative dimer-dimer motion, and $f(p_0)$ is the dimer-dimer scattering amplitude.

{ In order to understand the behavior of $F_{\1\2}(p_2,p_4)$ corresponding to the asymptote (\ref{PsiAsymp}) we multiply Eq.~(\ref{PsiAsymp}) by $\delta(x_{13})$ and Fourier transform it arriving at ${\cal F}_{13}(p_2,p_4;Q)=2\pi \delta(Q) F_{\1\2}(p_2,p_4)$ with $F_{\1\2}(p_2,p_4)\propto \tilde \phi_0[(p_2-p_4)/2] \tilde \chi(-p_2-p_4)$, where $\tilde\phi_0$ and $\tilde\chi$ are Fourier transforms of $\phi_0$ and $\chi$, respectively.} Accordingly, $F_{\1\2}(p_2,p_4)$ has a singularity at $|p_2+p_4|=p_0$, close to which it behaves as
\begin{equation}\label{Singularity}
F_{\1\2}(p_2,p_4)\propto \frac{2\pi \delta(|P|-p_0)-4ip_0f(p_0)/(P^2-p_0^2-i0)}{p^2+1/a_{\1\2}^2}, 
\end{equation}
where we have introduced the pair center-of-mass and relative-coordinate representation, $P=p_2+p_4$ and $p=(p_2-p_4)/2$. Motivated by Eq.~(\ref{Singularity}) we make the substitution
\begin{equation}\label{fthroughG}
F_{\uparrow \downarrow}(p_2,p_4) = \frac{2\pi \delta(|P|-p_{0})-4G(P,p)/(P^2 - p_{0}^2)}{p^2 + 1/a_{\1\2}^2}
\end{equation}
in the STM equations and obtain an inhomogeneous set of equations for $G$, $F_{\1\1}$, and $F_{\2\2}$. For numerical convenience we restrict ourselves to real-valued functions and understand integration of terms proportional to $1/(P^2-p_0^2)$ in the principal-value sense. Once $G$ is calculated, $f$ can be deduced by comparing Eqs.~(\ref{Singularity}) and (\ref{fthroughG}) and by using the convention $1/(P-p_0-i0)=1/(P-p_0)+\pi i \delta (P-p_0)$. We arrive at the identification
\begin{equation}\label{ScatAmplthroughG}
f(p_0)=-\frac{1}{1-ip_0/G(p_0)},
\end{equation} 
where $G(p_0)=G(p_0,p)$ does not depend on $p$. The value $G(p_0)$ thus determines the dimer-dimer scattering amplitude and, in particular, the dimer-dimer effective-range expansion through $G(p_0)=-1/a_{\rm dd}+r_e p_0^2/2+...$, where $a_{\rm dd}$ and $r_e$ are, respectively, the dimer-dimer scattering length and effective range. The dimer-dimer coupling constant is defined as $g_{\rm dd}=-1/a_{\rm dd}=G(0)$. More generally, one can think of $G(p_0)$ as the energy-dependent coupling constant. 

\begin{center}
\begin{figure}[ht]
\vskip 0 pt \includegraphics[clip,width=0.9\columnwidth]{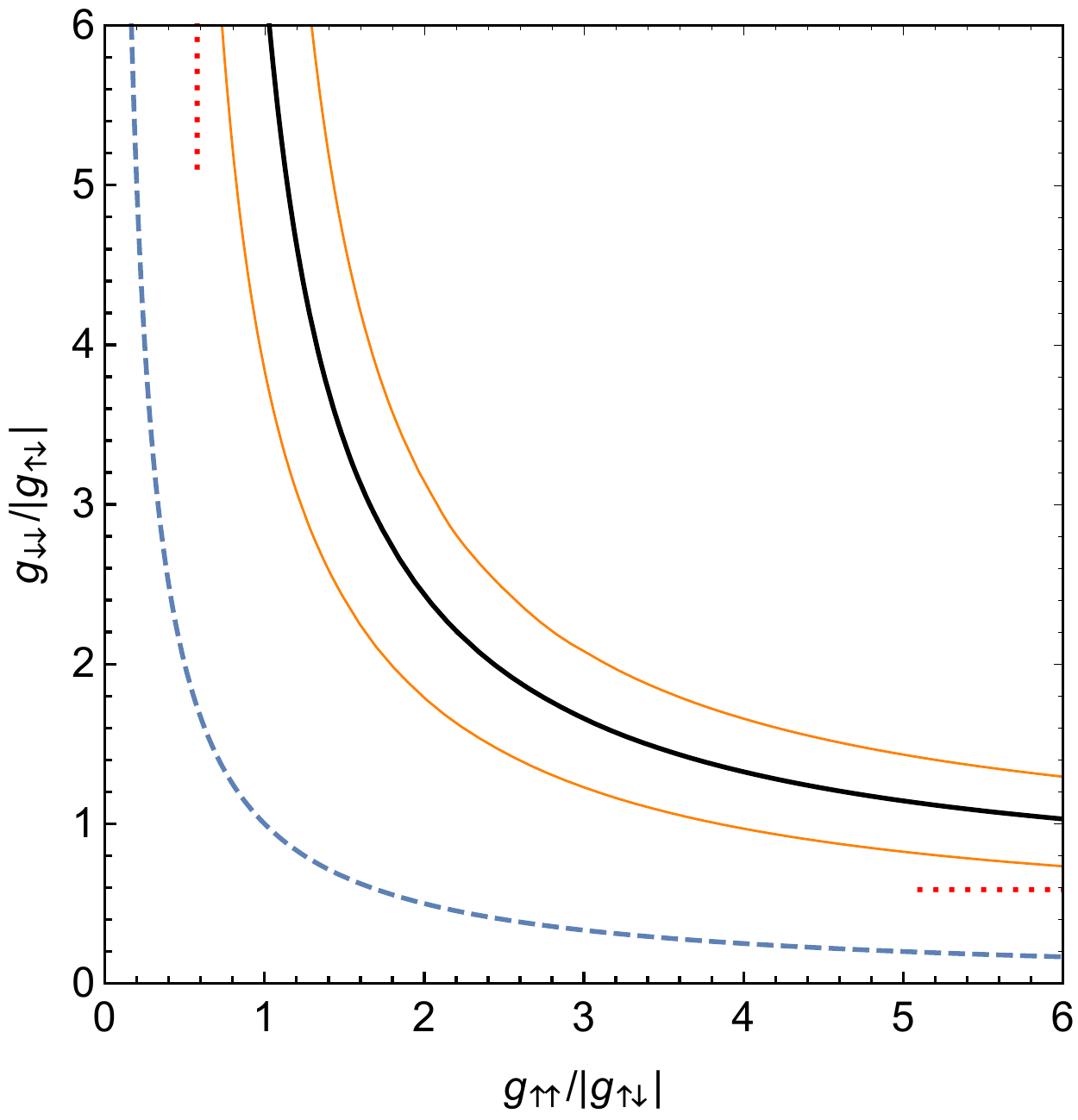}
\caption{
Zero crossing for the interaction between two $\1\2$ dimers (thick black). The dotted red horizontal and vertical lines are the corresponding asymptotes for $g_{\1\1}=\infty$ and $g_{\2\2}=\infty$, respectively. The thin orange curves indicate parameters where $g_{\rm dd}/|g_{\1\2}|=0.1$ (upper curve) and $-0.1$ (lower curve). The dashed blue curve is the mean-field collapse boundary $g_{\1\2}=-\sqrt{g_{\1\1}g_{\2\2}}$. 
}
\label{Fig:ZeroCrossing}
\end{figure}
\par\end{center}

The solid black curve in Fig.~\ref{Fig:ZeroCrossing} corresponds to the values of $g_{\1\1}/|g_{\1\2}|$ and $g_{\2\2}/|g_{\1\2}|$ where $g_{\rm dd}$ vanishes and $a_{\rm dd}$ diverges. This curve is obviously symmetric with respect to the interchange of $g_{\1\1}$ and $g_{\2\2}$. For infinite $g_{\1\1}$ the dimer-dimer zero crossing is located at { $g_{\2\2}/|g_{\1\2}|=0.575(3)$}. This asymptote is indicated by the horizontal dotted red line. The vertical one is its symmetric analog. The upper and lower thin orange curves correspond, respectively, to $g_{\rm dd}/|g_{\1\2}|=0.1$ (repulsion) and $-0.1$ (attraction). The dashed blue curve represents the collapse boundary $g_{\1\1}g_{\2\2}=g_{\1\2}^2$. 

We have also calculated the dimer-dimer effective range $r_e$ along the zero-crossing line. We find that $r_e/|a_{\1\2}|$ does not change much on the scale of Fig.~\ref{Fig:ZeroCrossing}. This ratio approximately equals $1.25$ for $g_{\1\1}=g_{\2\2}$ and increases to $1.3$ as one reaches the point where $g_{\sigma\sigma}/|g_{\1\2}|=6$. We see that $r_e$ is on the order of the dimer size. One can thus think of the dimer-dimer interaction in terms of an effective potential with the range $\sim r_e\sim a_{\1\2}$ and competing attractive and repulsive parts.

\section{Dilute liquid of dimers}\label{Sec:Liquid}

In this section we argue that sufficiently close to the dimer-dimer zero-crossing line, on its attractive side, many dimers form a dilute dimerized liquid. The liquid state is a result of a competition between two- and three-dimer forces as predicted by Bulgac \cite{Bulgac}. In the one-dimensional case that we consider the three-body scattering is kinematically equivalent to the two-dimensional two-body scattering and the corresponding mean-field energy shift depends logarithmically on the energy itself (see, for example, \cite{Petrov3body,Nishida}). This logarithmic running makes the mean-field description of the system slightly more complicated than in the three-dimensional case discussed in \cite{Bulgac}. On the other hand, it also allows us to make quantitative predictions of liquid properties without actually solving the three-dimer problem. In our analysis we will use analogies with the well-studied problem of two-dimensional two-body-interacting bosons.

Consider $N_{\rm d}>2$ dimers close to the dimer-dimer zero crossing in the attractive regime where $a_{\rm dd}\gg a_{\1\2}\sim r_e$. To the zeroth order in the dimer size one can think of the dimers as point-like particles neglecting their composite nature. As follows from Sec.~\ref{Sec:Trimer}, trimers can be excluded from this picture since they are not sufficiently deeply bound and we consider the population-balanced case. Thus, to the leading order, we deal with a gas of $N_{\rm d}$ attractive point-like bosons, the ground state of which is a soliton with the energy \cite{McGuireBosons}
\begin{equation}\label{McGuire}
E_{N_d}=-g_{\rm dd}^2 N_{\rm d}(N_{\rm d}^2-1)/12
\end{equation} 
and size $L\sim 1/\sqrt{\epsilon}\sim a_{\rm dd}/N_{\rm d}$, where $\epsilon \sim E_{N_d}/N_d$ is the energy per dimer and we do not count the dimer binding energies. The central density of dimers diverges with increasing $N_{\rm d}$ (keeping $a_{\rm dd}$ fixed). This is obviously an artefact of the point-like approximation. That the system does not collapse can be shown by contradiction. Indeed, if the average distance between dimers becomes smaller than their size, the mixture enters into the mean-field ``atomic'' regime where it should be mechanically stable since we are above the mean-field collapse line $g_{\1\1}g_{\2\2}=g_{\1\2}^2$ \cite{PethickSmith}. 

We will now show that a repulsive three-dimer interaction stops the grows of the dimer density at a much lower value $n_{\rm d}\ll 1/a_{\1\2}$. Before we discuss the three-dimer interaction energy shift let us give general considerations on the three-body scattering in one dimension \cite{Petrov3body,Nishida}. After separating the center-of-mass motion the configurational space of three dimers is a two-dimensional plane parameterized by the hyperradius $\rho=\sqrt{2/3}\sqrt{x_{12}^2+x_{13}^2+x_{23}^2}$ and hyperangle $\phi = \arcsin(x_{12}/\rho)$, where $x_{ij}$ is the distance between dimers $i$ and $j$. The three-dimer interaction is an effective potential originating from virtual excitations of internal and external degrees of freedom of three colliding dimers (pair-wise dimer-dimer processes are excluded to avoid double counting). This potential is thus physically localized at $\rho\sim a_{\1\2}$ and is characterized by the scattering length $a_3>0$ defined as the position of the (extrapolated) node of the zero-energy three-body wave function $\propto \ln(\rho/a_3)$. An important consequence of this hyper-two-dimensional kinematics is that at low energies the three-body interaction becomes repulsive, equivalent to a hard-wall constraint at $\rho=a_3$. Even without solving the three-dimer problem one can assume that for small $a_{\1\2}/a_{\rm dd}$  the scattering length $a_3$ can be approximated by its value at $a_{\rm dd}=\infty$ and that it is of the same order of magnitude as the dimer size $a_{\1\2}$ (we will return to this point in the next section). 

In order to proceed to the many-body problem we assume that the state of the system is homogeneous in the thermodynamic limit and that it is susceptible to the mean-field treatment. The corresponding applicability condition requires that the interaction-induced chemical potential $\mu$ be much smaller than the quantity $n_{\rm d}^2$, comparable to the chemical potential in the strongly interacting Tonks-Girardeau regime. The inequality $|\mu|\ll n_{\rm d}^2$ also means that there is a macroscopic number of dimers per healing length (allowing for the classical description), where the healing length is $\sim 1/\sqrt{|\mu|}$. Here we require that the mean-field condition be satisfied separately for the two- and three-dimer interaction parts. In particular, for the two-dimer part we need $a_{\rm dd}n_{\rm d} \gg 1$. 

The treatment of the three-dimer interaction proceeds in the same manner as for the short-range two-body interaction in the two-dimensional case (see, for example, \cite{Schick,Popov,MoraCastin}). In one way or another, these approaches consist of replacing the short-range (in our case three-dimer) potential by an effective potential with the same scattering length but with the range larger than the mean interparticle distance and smaller than the healing length. This effective potential then looks short ranged at relevant momenta $\sim \sqrt{|\mu|}$ and, on the other hand, it is sufficiently weak for the applicability of the Born-series expansion. In our case, the three-dimer effective potential can be taken as a constant in momentum space \cite{Nishida},
\begin{equation}\label{g3}
g_3 = \frac{\sqrt{3}\pi}{2\ln(2e^{-\gamma}/a_3\kappa)},
\end{equation}
where $\gamma\approx 0.5772$ is the Euler constant and the overall numerical coefficient (different from the two-dimensional two-body scattering case) is related to the Jacobian of the transformation from the coordinates $x_{12}$ and $x_{23}$ to the two-dimensional hyperradius-hyperangle pair \cite{Petrov3body}. The potential is assumed to vanish above the (hyper)momentum cut-off $\kappa$, which satisfies $n_{\rm d} \ll \kappa \ll \sqrt{|\mu|}$. The three-body energy shift per dimer then equals $g_3 n_{\rm d}^2/6$ and the mean-field applicability condition is $g_3\ll 1$, which is satisfied, in particular, if $a_3$ is exponentially smaller than the mean inter-dimer separation. Note that in this case $g_3>0$. 

For negative $g_{\rm dd}$ and positive $g_3$ the energy per dimer
\begin{equation}\label{EnPerDim}
\epsilon=g_{\rm dd}n_{\rm d}/2+g_3 n_{\rm d}^2/6
\end{equation}
has a minimum at a finite saturation density \cite{Bulgac} given by 
\begin{equation}\label{nd}
n_{\rm d}=-3g_{\rm dd}/2g_3.
\end{equation}
For this density $\epsilon=\mu=-(3/8)g_{\rm dd}^2/g_3$ and it is easy to see that the two- and three-body mean-field applicability conditions reduce to $g_3\ll 1$.

The exact value of the cut-off momentum $\kappa$ is, in fact, not important if one sticks to the leading order in $g_3$. At this level of approximation the three cut-off values $\kappa=n_{\rm d}\sim 1/a_{\rm dd}g_3$, $\kappa=\sqrt{\mu}\sim 1/a_{\rm dd}\sqrt{g_3}$, and $\kappa = 1/a_{\rm dd}$ lead to the same result since they differ only by a power of $g_3$ rather than exponentially. Indeed, by substituting these cut-off momenta into Eq.~(\ref{g3}) one obtains three values of $g_3$ different from each other by $\sim g_3^2 \ln g_3\ll g_3$. We thus take $\kappa=1/a_{\rm dd}$. Similarly, we neglect other numerical factors under the logarithm and set $a_3=a_{\1\2}$ in Eq.~(\ref{g3}). This leads to the explicit expressions $g_3=\sqrt{3}\pi/2\ln(a_{\rm dd}/a_{\1\2})\ll 1$,
\begin{equation}\label{ndfin}
n_{\rm d}=(\sqrt{3}/\pi a_{\rm dd})\ln(a_{\rm dd}/a_{\1\2}),
\end{equation}
and
\begin{equation}\label{EnDens}
\mu=\epsilon=-(\sqrt{3}/4\pi a_{\rm dd}^2)\ln(a_{\rm dd}/a_{\1\2}).
\end{equation}

With increasing $N_{\rm d}$ the peaked soliton solution described by Eq.~(\ref{McGuire}) transforms into a liquid-like droplet characterized by an approximately constant bulk density $n_{\rm d}$ given by Eq.~(\ref{ndfin}). By comparing densities or energies per dimer in these two limits one can see that the soliton-droplet crossover happens at $N_{\rm d}\sim \sqrt{\ln(a_{\rm dd}/a_{\1\2})}$.

\section{Discussion and outlook}\label{Sec:Discussion}

The perturbative expansion in powers of $g_3\ll 1$ can be continued beyond the mean-field term. The next order requires the application of the Popov theory \cite{Popov} and a more precise knowledge of $a_3$. Note that dimer-dimer effective-range effects are beyond this power-law expansion. The effective-range energy correction per dimer scales as $r_e \epsilon n_{\rm d}$ and is thus smaller than $\epsilon$ by $r_e n_{\rm d} \sim g_3^{-1} e^{-\sqrt{3}\pi/2g_3}$ which is smaller than any power of $g_3$. The mean-field and beyond-mean-field treatment of effective-range effects for one-dimensional bosons has been discussed in Refs.~\cite{Sgarlata2015,Cappellaro}.

In Sec.~\ref{Sec:Liquid} we have substituted $a_{\1\2}$ for $a_3$ since distinguishing these two quantities is exceeding the accuracy of the leading-order calculation in the low-energy dilute regime defined by $a_{\1\2}\sim a_3 \ll 1/n_{\rm d}$ or, more precisely, by $1/\ln (1/a_{\1\2}n_{\rm d}) \sim 1/\ln(1/a_3 n_{\rm d}) \ll 1$. An interesting alternative appears in the regime
\begin{equation}\label{highenergy}
a_{\1\2} \ll 1/n_{\rm d} \ll a_3
\end{equation}
corresponding to a weak three-body attraction studied by Sekino and Nishida \cite{Nishida}. More precisely, they find that one-dimensional bosons with a pure three-body zero-range attraction form solitons with binding energies exponentially increasing and sizes exponentially decreasing with $N_{\rm d}$, similar to solitons of two-body-interacting two-dimensional bosons discovered by Hammer and Son \cite{HammerSon}. The relevance of the Sekino-Nishida states for our mixture depends on the exact solution of the three-dimer problem. We distinguish two possibilities: (1) There is no three-dimer bound state. This typically corresponds to $a_3$ being comparable to or smaller than the dimer size. In this case, our dilute liquid is a stable state. (2) $a_3$ is larger than the dimer size and the three-dimer bound state exists on or even above the dimer-dimer zero crossing. In this case our dilute solution still exists but becomes metastable with respect to the formation of clusters of the higher-density Sekino-Nishida phase. The fate of this phase in this case is an interesting problem by itself since the unlimited grows of density with $N_{\rm d}$ would eventually contradict the first inequality of (\ref{highenergy}). In any case, this discussion motivates solving the three-dimer ($\1\1\1\2\2\2$) problem, calculating $a_3$, and looking for eventual three-dimer bound states. An interesting possibility to check is whether there is a three-dimer zero-crossing point on the dimer-dimer zero-crossing curve.  
 
It is tempting to speculate on the behavior of our dilute dimerized liquid as one moves from the dimer-dimer zero crossing towards the mean-field collapse curve in Fig.~\ref{Fig:ZeroCrossing}. The liquid phase just above this curve has been studied in Ref.~\cite{PetrovAstrakharchik}. Pairing correlations have not been discussed for this ``atomic'' liquid, but they seem to be irrelevant for its self-trapping character. Moreover, Ref.~\cite{PetrovAstrakharchik} suggests that for $g_{\1\1}\neq g_{\2\2}$ the liquid is density imbalanced, $n_\1/n_\2=\sqrt{g_{\2\2}/g_{\1\1}}$. Hence, for example, for $g_{\1\1}>g_{\2\2}$, a dimerized liquid droplet does not adiabatically connect to the atomic one as it should somehow get rid of the $\1$ component. We thus expect more details to appear in the many-body analog of the diagram in Fig.~\ref{Fig:ZeroCrossing}, at least, outside of the diagonal $g_{\1\1}=g_{\2\2}$. Note that this region can be investigated experimentally in the two-component mixture of $^{39}$K studied in Refs.~\cite{Tarruell1,Fattori,Tarruell2}. 
These speculations bring up a potentially interesting few-body problem in the $\1\1\1\2\2$ configuration{, which can be thought of as two dimers and an atom. As we have shown (see Fig.~\ref{Fig:Trimer}), the atom always binds to a dimer. In addition, it can hop from one dimer to the other thus mediating an exchange attraction between them. Therefore, even above the dimer-dimer zero crossing this attraction can overcome the dimer-dimer repulsion and bind the system into a pentamer state.} More generally, this scenario suggests that the liquid phase in the population-imbalanced configuration may extend above the zero-crossing curve in Fig.~\ref{Fig:ZeroCrossing}.

Finally, there appears a fascinating possibility of observing self trapping in a one-dimensional Fermi-Bose mixture with interspecies attraction and Bose-Bose repulsion. We predict that fermionic Fermi-Bose dimers in this case bind for $g_{\rm BB}<0.575 |g_{\rm FB}|$ (assuming equal masses) and can thus form a dilute dimerized liquid. This effect can be studied, in particular, in the $^{40}$K-$^{41}$K mixture by utilizing the wide interspecies Feshbach resonance at 540 G \cite{Zwierlein}. Very recently, Pan and coworkers \cite{Cui} have discussed a one-dimensional (atomic) Fermi gas near a $p$-wave resonance and argued that in the collapse regime (equivalent to our $g_{\rm dd}<0$) the system can be stabilized by effective-range effects. No three-atom interaction is included in their model.

\section*{ACKNOWLEDGMENT}

We thank G. Astrakharchik and L. Tarruell for useful discussions. This research was supported by the European Research Council (FR7/2007-2013 Grant Agreement No. 341197).

\end{document}